\documentclass[aps,pra,superscriptaddress,10pt]{revtex4-2}

\usepackage[T1]{fontenc}
\usepackage{dsfont}
\usepackage{amsmath}
\usepackage{bm}
\usepackage{amssymb}
\usepackage{graphicx}
\usepackage{subfigure}
\usepackage{comment}
\usepackage{xcolor}
\usepackage{oplotsymbl}
\usepackage{lmodern}
\usepackage{empheq}
\usepackage[colorlinks=true,urlcolor=blue,citecolor=blue,linkcolor=blue]{hyperref}
\usepackage{float}	
\newcommand{\ie}{\emph{i.e.}}
\newcommand{\Gr}{\text{Gr}}

\newcommand{\ket}[1]{| #1 \rangle}
\newcommand{\braket}[2]{\langle #1 | #2 \rangle}
\newcommand{\pyr}[1]{\ket{\psi_{\triangle}^{(#1)}}}
\newcommand{\apyr}[1]{\ket{\psi_{\bigtriangledown}^{(#1)}}}
\newcommand{\bpy}[2]{\ket{\psi_{\diamond}^{(#1,#2)}}}

\begin{document}

\title{Anticoherent $k$-planes and coding techniques for a 3-qubit scheme of universal quantum computing}

\author{L.{} Arag\'on-Mu\~noz$^{*}$, C.{} Chryssomalakos$^{\dagger}$, A.{} G.{} Flores-Delgado$^{\ddagger}$, V.{} Rascón-Barajas$^{\S}$, and I.{} V\'azquez Mota$^{\P}$}

\address{Instituto de Ciencias Nucleares, Universidad Nacional Aut\'onoma de M\'exico,\\
	P.O. Box 70-543, 04510 Ciudad de M\'exico, M\'exico\\
	$^{*}$luis.aragon@correo.nucleares.unam.mx\\
	$^{\dagger}$chryss@nucleares.unam.mx\\
	$^{\ddagger}$ana.flores@correo.nucleares.unam.mx\\
	$^{\S}$valentina.rascon@correo.nucleares.unam.mx\\
	$^{\P}$igor.vazquez@correo.nucleares.unam.mx}

\begin{abstract}
	
	Toponomic quantum computing (TQC) employs rotation sequences of anticoherent $k$-planes to construct noise-tolerant quantum gates. In this work, we demonstrate the implementation of generalized Toffoli gates, using $k$-planes of spin systems with $s \geq k + 1$, and of the Hadamard gate for a 3-qubit system, using a spin $s \!= \! 15$ 8-plane. We propose a universal quantum computing scheme for 3-qubit systems (via Hadamard + Toffoli gates) based on coding techniques. A key advantage of this construction is its inherent robustness against noise: apart from reparametrization invariance, our scheme is characterized by immunity to arbitrarily large deformations of the path in (rotational) parameter space.

\end{abstract}
\date{\today}
\maketitle

\section{Introduction}

Developing quantum computing protocols that are robust against noise is one of the main challenges in achieving reliable and efficient quantum computation~\cite{Zan.Ras:99, Sol.Zan.Zan:04}. Various types of noise affect quantum systems~\cite{Geo.Ema.Zul:21,Kni.Laf.Zur:98}, including decoherence due to interactions with the environment, control errors, and imperfections in gate implementation. In this work, we show how to implement various quantum gates in the framework of holonomic quantum computing (HQC), with an extra topological ingredient that eliminates the effects of, \emph{e.g.}, stray magnetic fields.

The TQC scheme introduces a new feature to HQC~\cite{Chr.Han.Guz.Ser:22}: it requires that the $k$-plane used in the quantum protocol satisfy a condition known as \textit{anticoherence}. When such subspaces possess a rotational symmetry, the resulting logical gate, which is constructed through a rotation sequence, starting at the identity and ending at the symmetry rotation, is not only invariant under reparametrizations of the path, but also under (arbitrarily large) deformations of the path traced in $\text{SO}(3)$. Along with the obvious advantages of this approach, one encounters two main challenges: (i) finding the particular $k$-planes the scheme calls for and (ii) identifying the rotational symmetries that enable the implementation of specific quantum gates.

For the purposes of this paper, we find the ``hand-made'' approach to the construction of anticoherent $k$-planes put forth in Ref.~\cite{Chr.Han.Guz.Ser:22} quite useful. As described in detail in Section 3, this method relies on a particular choice of quantum states, the Majorana constellations%
\footnote{%
	The Majorana constellation of a spin-$s$ state is a geometric representation of the state via $2s$ points on the unit sphere~\cite{Maj.Rep}.%
}
of which are analyzed to discern specific rotational symmetries, which, in turn, imply rotational symmetries of the $k$-plane as a whole%
\footnote{%
	A more formal approach, developed in Ref.~ \cite{Chr.Han.Guz.Ser:22}, relies on the plane's \emph{multiconstellation}, a generalization of the Majorana constellation for $k$-planes --- see Ref.~ \cite{Chr.Guz.Han.Ser:21}.%
} --- 
we construct, in this way, generalized Toffoli gates, using $k$-planes for any spin $s\geq k+1$. In Section 4, we present the main result of the paper: a universal quantum computing scheme using two distinct 8-planes in spin 15. This is achieved through a specific technique known as \textit{coding}, describing the implementation of the Hadamard and Toffoli gates, which, together, form a universal set of quantum gates~\cite{Aha:03}.

In what follows, we introduce the necessary geometric tools and set up the underlying formal framework.

\section{Mathematical preliminaries}
\label{sec:mat_pre}

A cyclic and adiabatic evolution of a quantum state in the projective space gives rise to a geometric phase factor~\cite{Ber:84}, which is invariant under reparametrizations of the evolution curve. 
A non-abelian generalization of this concept was introduced in Ref.~\cite{Wil.Zee:84} and Ref.~\cite{Sjo:15}, associated with the evolution of a $k$-dimensional subspace of the Hilbert space $\mathcal{H}\simeq \mathbb{C}^{N}$ ($N=2s+1$). In this case, the geometric ``phase'' factor takes the form of a $k \times k$ unitary matrix. This generalization requires the formulation of the problem in the Grassmannian $\Gr(k, N)$, i.e., the space of $k$-planes passing through the origin in $\mathbb{C}^{N}$. We identify a $k$-plane $\Pi$ (\ie, a point in $\Gr(k, N)$) by an orthonormal basis that spans it --- this basis will be associated with the input qubits when implementing the quantum computing protocol.

The HQC method for implementing a quantum gate considers a closed curve in $\text{\Gr}(k,N)$, $\Pi(t) = \textup{span}\{\ket{\psi_i(t)}\}$, $t \in [0,1]$, with $\Pi(0) = \Pi(1)$ and the $\ket{\psi_i(t)}$ orthonormal for all $t$, but otherwise arbitrary --- in particular, we do not assume that $\ket{\psi_i(1)}=\ket{\psi_i(0)}$ and define the \emph{overlap} matrix  $W$ by $\ket{\psi_i(1)}=\sum_{j} \ket{\psi_j(0)}W_{ji}$. To this closed curve, one associates a unitary $k \times k$ matrix $U$, given by
\begin{equation}
	\label{eq:Udyn}
	U
	=
	W \textup{Pexp}\left(-\int_{0}^{1}dt\,\mathcal{A}(t)\right),
	\quad
	\left(
	\mathcal{A}_{ij}(t)=\braket{\psi_{i}(t)}{\dot{\psi}_{j}(t)}
	\right)
	\, ,
\end{equation}
where $\text{Pexp}$ denotes a path-ordered exponential and $\mathcal{A}$ is known as the Wilczek-Zee (WZ) connection~\cite{Wil.Zee:84}. This association has both a mathematical and a physical motivation. From the mathematical point of view, the above unitary matrix gives the holonomy of the horizontal lift of the curve $\Pi(t)$ in the frame bundle above $\Gr(k,N)$, where horizontality amounts to the derivatives of the frame vectors being orthogonal to $\Pi(t)$. From the physical point of view, one obtains exactly this kind of evolution, if all vectors in $\Pi(t)$ are energy-degenerate, and the corresponding Schr\"odinger evolution is adiabatic. Then, any ket initially in $\Pi(0)$, stays in $\Pi(t)$ for all $t$, and is transformed by the above $U$ at the end of the cycle. It is worth pointing out that the above condition of adiabaticity, which might well clash with the demands of quantum computing, can be lifted~\cite{Ana:88}, and even the closure of the curve in the Grassmannian is optional~\cite{Muk.Sim:93}.

In TQC, additional ingredients  increase robustness against noise in the implementation of the logic gate. The method, presented in Ref.~\cite{Chr.Han.Guz.Ser:22}, is described below:

\begin{itemize}
	\item 
	A spin-$s$ anticoherent $k$-plane $\Pi=\textup{span}\{\ket{\psi_{i}}\}$ is identified --- this means that any basis $\{\ket{\psi_i}\}$ in $\Pi$ satisfies
	\begin{equation}
		\braket{\psi_{i}}{S_{A}|\psi_{j}}
		=
		0
		\, ,
		\quad
		A=x,y,z, 
		\quad
		i=1,2,\ldots,k
		\, ,
		\label{eq:1_AC}
	\end{equation} 
	where $S_{x}$, $S_{y}$, $S_{z}$ are the cartesian components of the spin $s$ angular momentum operator.
	\item
	Additionally, $\Pi$ is required to have a rotational symmetry $R_{\mathbf{n}_1} \in \textup{SO}(3)$,
	
	\[
	R_{\mathbf{n}_{1}}\bigl( \Pi\bigr)
	\equiv
	\textup{span}\{D^{(s)}(\widetilde{R}_{\mathbf{n}_{1}})\ket{\psi_{i}}\}
	=
	\Pi
	\, ,
	\]
	where $\mathbf{n}_{1}$ points along the rotation axis and has modulus equal to the rotation angle, $\widetilde{R}_{\mathbf{n}_{1}}$ is the lift%
	\footnote{%
		$\textup{SU}(2)$ being a double cover of $\textup{SO}(3)$, every rotation $R_{\mathbf{n}}\in \textup{SO}(3)$ corresponds to two distinct elements in $\textup{SU}(2)$ that differ by a sign. In what follows we map the identity element of $\textup{SO}(3)$ to the identity element in $\textup{SU}(2)$, so that any $\text{SO}(3)$ curve that starts at the identity, has a unique lift in $\text{SU}(2)$.%
	} 
	of $R_{\mathbf{n}_{1}}$ to $\textup{SU}(2)$ and $D^{(s)}(\widetilde{R}_{\mathbf{n}_{1}})$ is its spin-$s$ irreducible representation.	
	\item
	A smooth curve in $\textup{SO}(3)$ is chosen that joins the identity with $R_{\mathbf{n}_{1}}$,
	\[
	t\in[0,1]\mapsto R_{\mathbf{n}(t)},\quad \mathbf{n}(0)
	=
	\mathbf{0},\quad \mathbf{n}(1)=\mathbf{n}_{1}
	\, ,
	\]
	and a closed curve $\Pi(t)$ in $\Gr(k,N)$ is obtained by rotating $\Pi$ by $R_{\mathbf{n}(t)}$,
	\[
	\Pi(t)
	=
	R_{\mathbf{n}(t)}\bigl( \Pi\bigr)
	=
	\textup{span}\Bigl\{D^{(s)}(\widetilde{R}_{\mathbf{n}(t)})\ket{\psi_{i}}\Bigr\}
	\, .
	\]
	Note that $\Pi(t)$ remains, at all times, anticoherent as a consequence of the rotational invariance of the anticoherence conditions (\ref{eq:1_AC}).
\end{itemize}
It is precisely because of the anticoherence of $\Pi(t)$, and the fact that the derivative of $\ket{\psi_{j}(t)}=D^{(s)}(\widetilde{R}_{\mathbf{n}(t)})\ket{\psi_{j}}$ is of the form $m^{A}(t) S_{A}\ket{\psi_{j}(t)}$ (for some vector $\mathbf{m}(t)$), that the WZ connection vanishes along the entire curve,
\[\mathcal{A}_{ij}(t)
=
\langle \psi_{i}(t)|\dot{\psi_{j}}(t)\rangle
=
m^{A}(t)\langle \psi_{i}(t)|S_{A}|\psi_{j}(t)\rangle=0\, ,\]
which reduces the holonomy $U$ to the matrix $W$:
\begin{equation}
	U_{ij}
	=
	\langle \psi_{i}(0)|\psi_{j}(1)\rangle
	=
	\langle \psi_{i}|D^{(s)}(\widetilde{R}_{\mathbf{n}_{1}})|\psi_{j}\rangle\, .
	\label{eq:Utop}
\end{equation}
The noise tolerance of $U$ stems from its topological character: it is not the specific shape of the curve $t\mapsto \Pi(t)$ that determines $U$, but rather its homotopy class: in the presence of a rotational discrete symmetry group $G$, the orbit $\mathcal{O}(\Pi)$ of $\Pi$ under rotations is diffeomorphic to $SO(3)/G$, the fundamental group of which is isomorphic to $G$~\cite{Mer:79} --- for example, a tetrahedral state has a rotational  orbit with twelve distinct homotopy classes (in 1-to-1 correspondence with the twelve elements of the rotational tetrahedral group), curves in each class being continuously deformable into each other, but not into curves from a different class. Thus, a closed curve on $\mathcal{O}(\Pi)$ has, in general, plenty of opportunities to get tangled up in the intricate underlying topology, and the corresponding holonomy $U$ only depends on the class it is in.
\section{A systematic method to generate anticoherent $k$-planes}
In recent years, spin-$s$ anticoherent states (\emph{i.e.}, anticoherent 1-planes) have gained relevance because of their ``maximally non-classical'' status~\cite{Rud.Bur.Zyc:24}, which makes them indispensable in the study of purely quantum phenomena, such as entanglement~\cite{Bag.Bas.Mar:14}, and in quantum metrology~\cite{Chr.Her:17,Mar.Wei.Gir:20} --- as a consequence, they are, by now, well studied~\cite{Bag.Dam.Gir.Mar:15,Zim:06}. On the other hand, despite occasional mentions of anticoherent $k$-planes in the literature~\cite{Per.Pau:17,Ser.Mar.Chr:25}, our current knowledge on the subject is, at best, sketchy. Thus, we are forced to resort to \emph{ad hoc} recipes: in this section we use two well-known types of quantum states, pyramidal and bipyramidal~\cite{Bag.Dam.Gir.Mar:15}, to construct anticoherent $k$-planes, which are then subject to symmetry rotations to produce quantum gates in the TQC framework.
\subsection{Generalized Toffoli gates in pyramid-bypiramid $k$-planes}
\begin{figure}[H]	\centerline{\includegraphics[trim=1cm 1cm 1cm 1cm,clip,width=0.3\textwidth]{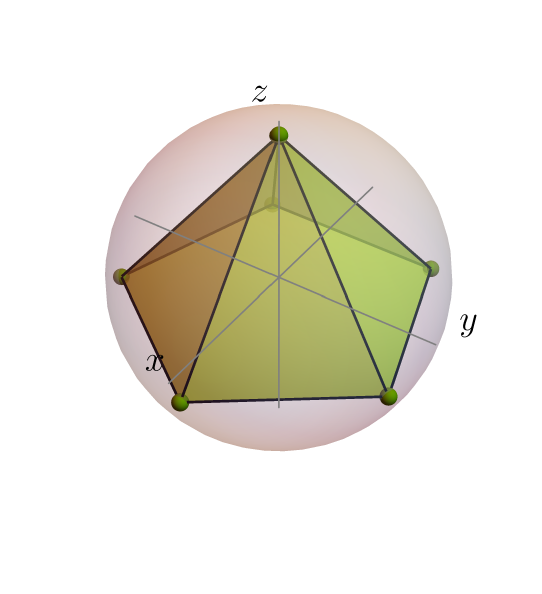}\includegraphics[trim=1cm 1cm 1cm 1cm,clip,width=0.3\textwidth]{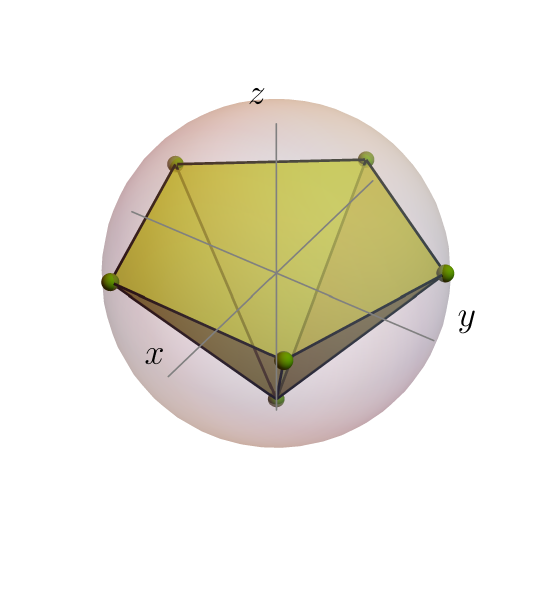}\includegraphics[trim=1cm 1cm 1cm 1cm,clip,width=0.3\textwidth]{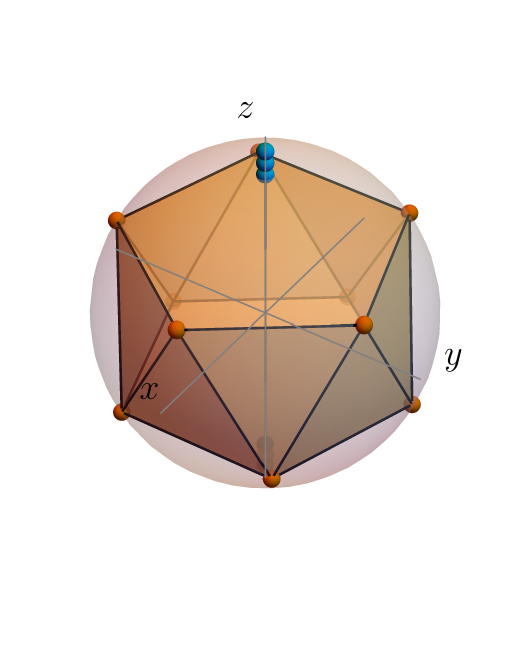}}
	\vspace*{8pt}
	\caption{Majorana constellations for $\ket{\psi^{(3)}_{\triangle}}$ (left) and for $\ket{\psi^{(3)}_{\bigtriangledown}}$ (center). The rightmost figure shows the multiconstellation (disregarding the spectator constellation) that represents the 2-plane  $\Pi_{\triangle}$. The orange points represent the principal constellation (corresponding to spin $s=5$) while the blue points represent the secondary constellation (corresponding to spin $s=3$).}
	\label{fig:pyrPlane}
\end{figure}

A spin-$s$ pyramidal state ($s \geq 2$) is of the form
\begin{equation}
	\pyr{s}
	=
	\sqrt{\frac{s-1}{2s-1}} \left(\ket{s,s} + \sqrt{\frac{s}{s-1}}\ket{s,-s+1}\right)\, ,
	\label{eq:pyramidal_state}
\end{equation}
where $\ket{s,m}$ is an element in the standard $(S^{2},S_{z})$ eigenbasis (the $m$-basis). These states are anticoherent~\cite{Bag.Mar:17}.
The Majorana constellations of pyramidal states consist of a star in the north pole, and the remaining $2s-1$ stars forming a regular polygon in a plane parallel to the equator. 
Associated with every integer-spin pyramidal state $\pyr{s}$, we define its ``partner'' $\apyr{s}$, which is the state with antipodal Majorana constellation (see Fig. \ref{fig:pyrPlane}),
\begin{equation}
	\apyr{s}
	=
	\sqrt{\displaystyle\frac{s-1}{2s-1}}\Bigl(\ket{s,-s}-\sqrt{\displaystyle\frac{s}{s-1}}\ket{s,s-1}\Bigr)
	\, ,
	\label{eq:pyramidal_state_2}	
\end{equation}
while for half-integer spins, the partner state $\ket{\psi^{(s)}_{\bigtriangledown}}$ is the antipodal, rotated around the $z$ axis by $\pi/(2s-1)$.

Each of the states $\pyr{s}$ and $\apyr{s}$ is anticoherent due to their rotational symmetries~\cite{Bag.Dam.Gir.Mar:15}, and the precise positioning of the base with respect to the apex. This, by itself, does not guarantee that $\Pi_{\triangle}$ also is, as linear combinations of anticoherent states are not necessarily anticoherent. It can be easily checked though that $\Pi_{\triangle}$ is indeed anticoherent.

For integer spin, a rotation by $\pi$ around the $y$ axis interchanges the basis elements but leaves the plane itself invariant. Therefore, a quantum NOT gate $\sigma_{x}$ is obtained by evaluating the holonomy of $\Pi_{\triangle}$, along the curve $R_{\mathbf{n}(t)}(\Pi_{\triangle})$ with $\mathbf{n}(1) = \pi \hat{y}$, in the basis $\{\pyr{s}, \apyr{s}\}$:
\begin{equation}
	U_{\textup{NOT}}
	=
	\left(
	\begin{array}{cc}
		0&1\\
		1&0
	\end{array}\right)=\sigma_{x}\, .
	\label{eq:U_not}
\end{equation} 

Another family of anticoherent states are the $m$-bipyramids: considering an integer spin $s$ and a non-negative eigenvalue $m$ of $S_z$, we define 
\begin{equation}
	\bpy{s}{m}
	=
	\left\{\begin{array}{lll}
		\displaystyle\frac{1}{\sqrt{2}}\Bigl(\ket{s,m}+\ket{s,-m}\Bigr),&\qquad& m\neq 0\, ,
		\\
		\\
		\ket{s,0},&&m=0\, .
	\end{array}\right.
	\label{eq:Byp}
\end{equation}
The anticoherence of the bipyramidal states is due to the ample spacing between their nonzero components, as the spin operators $S_A$ can only connect $S_z$-eigenstates with eigenvalues that differ at most by 1.
The Majorana constellation of $\bpy{s}{m}$ consists of $s-m$ stars at the north pole of the sphere, $2m$ stars distributed in a regular $2m$-polygon at the equator and the remaining $s-m$ stars at the south pole, thus forming a $2m$-gonal bipyramid (see figure \ref{fig:(5/2,m)-bipyramids}).

\begin{figure}[H]
	\centering
	\includegraphics[trim=1cm 1cm 1cm 1cm,clip,width=0.225\textwidth]{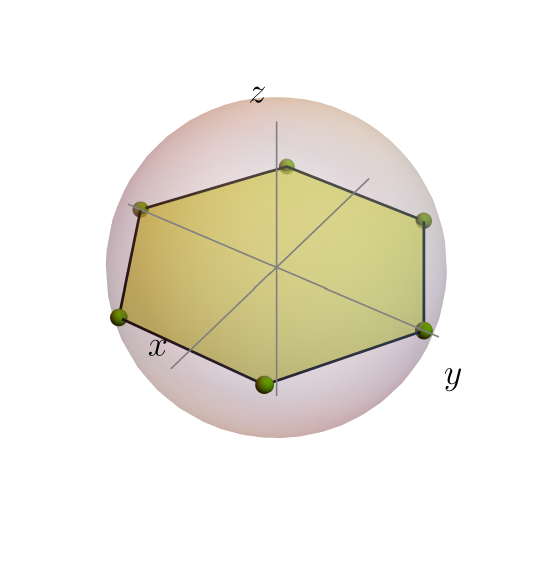}
	\includegraphics[trim=1cm 1cm 1cm 1cm,clip,width=0.225\textwidth]{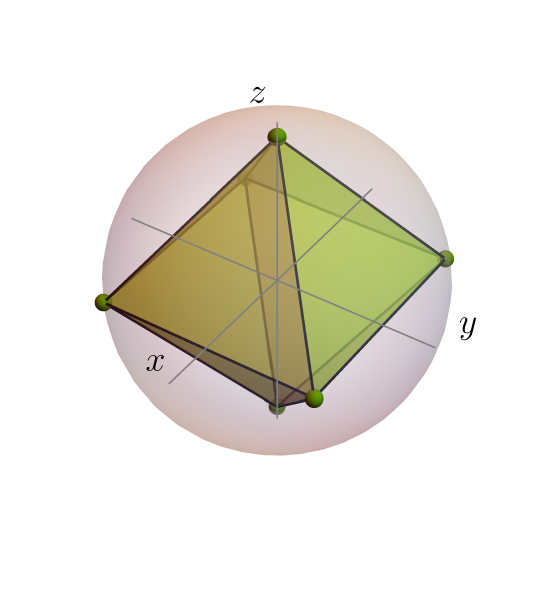}
	\includegraphics[trim=1cm 1cm 1cm 1cm,clip,width=0.225\textwidth]{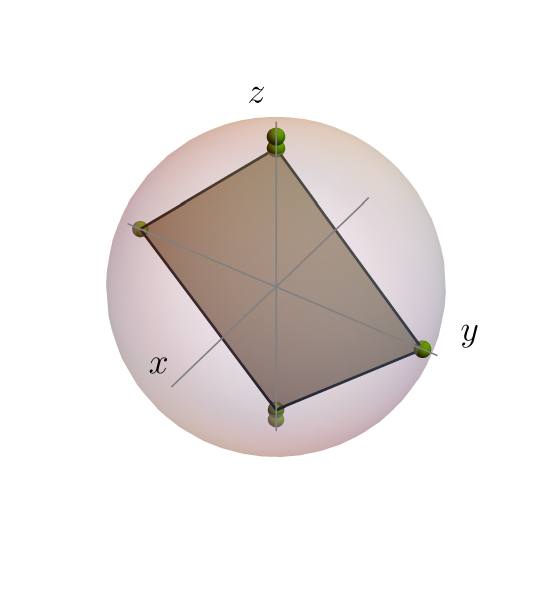}
	\includegraphics[trim=1cm 1cm 1cm 1cm,clip,width=0.225\textwidth]{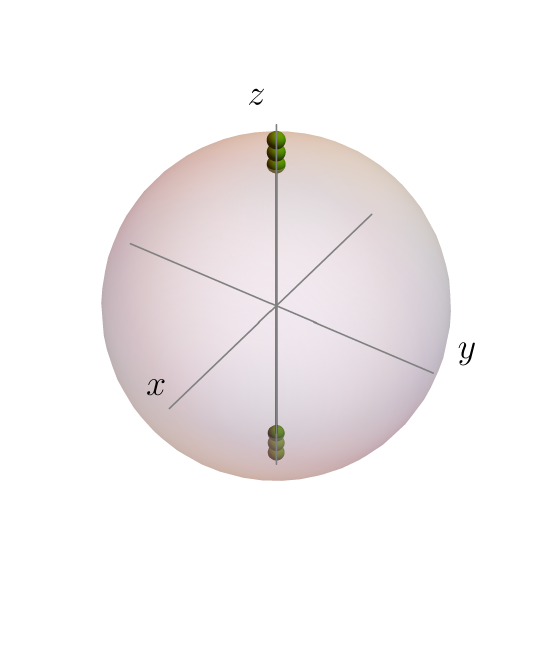}
	\caption{Majorana constellations of spin-3 bipyramidal states $\bpy{3}{m}$, for $m = 3, \, 2,\, 1,\, 0$ (left to right).}
	{\label{fig:(5/2,m)-bipyramids}}
\end{figure}
All bipyramidal Majorana constellations are invariant under the rotation $R_{\pi \hat{y}}$.  For $m$ even, the states in (\ref{eq:Byp}) acquire a global phase factor $(-1)^{s}$ upon rotation.
Therefore, for any $k \geq 4$ and $s\geq 2k-3$ integer and even, the two states $\pyr{s}$ and $\apyr{s}$, together with the $k-2$ bipyramidal states $\bpy{s}{2(j-1)}$  ($j =1,2,..,k-2$), span an anticoherent $k$-plane
\begin{align}
	\Pi_{\diamond}
	=
	\textup{span}\Bigl\{
	\bpy{s}{0},
	\bpy{s}{2},
	...,
	\bpy{s}{2(k-3)},
	\pyr{s},
	\apyr{s}\Bigr\}\, .
\end{align}
To see that $\Pi_{\diamond}$ is anticoherent, note that bipyramidal and pyramidal states used in the construction of the $k$-plane are known to be anticoherent, as previously proved and cited, respectively.
This means that the diagonal matrix elements of the spin operators $S_A$  in (\ref{eq:1_AC}) vanish. Moreover, the ample spacing of the nonzero components of the states involved guarantees that the off-diagonal elements of $S_A$ in (\ref{eq:1_AC}) also vanish.

The holonomy associated with the closed curve $R_{\mathbf{n}(t)}(\Pi_{\diamond})$ ($\mathbf{n}(1)=\pi \hat{y}$), in the chosen basis, corresponds to a generalized Toffoli gate with $k-1$ control qubits and one target,
\begin{equation}
	T
	=
	I_{k-2}\oplus \sigma_{x}
	\, .
	\label{eq:tof}
\end{equation}
\subsection{Toffoli and Hadamard gates in bypiramidal $8$-planes}
Returning to the bipyramidal states in (\ref{eq:Byp}), we note that the 8-plane 
\begin{equation}
	\label{eq:pi_1}
	\Pi_{1}
	=
	\textup{span}\left\{
	\bpy{15}{0},
	\bpy{15}{2},
	...,
	\bpy{15}{14}
	\right\}
\end{equation}
is anticoherent. In addition to the symmetry with respect to the rotation $R_{\pi \hat{y}}$, which in this plane yields the trivial holonomy $-I_{8}$ (with a phase factor due to the fact that $s$ is odd), we also find that $\Pi_{1}$ exhibits symmetry under rotations $R_{\frac{n\pi}{2} \hat{z}}$ for $n = 1, 2, 3$. This additional symmetry is a consequence of the fact that the number of vertices at the base of each bipyramid is a multiple of 4.
In particular, for $n=1$, the matrix in (\ref{eq:Utop}) is given by the direct sum of four copies of $\sigma_z$,
\begin{equation}
	U_{H}
	=
	\sigma_{z} \oplus \sigma_z \oplus \sigma_z \oplus \sigma_z = I_4 \otimes \sigma_z \equiv \sigma_z^{\oplus 4}
	\, .
	\label{eq:U_byp_1}
\end{equation}
This matrix coincides with the diagonal form of the Hadamard gate ($H$) applied to the third qubit of a tripartite system, given by
\begin{equation}
	H_3
	=
	I_{2} \otimes I_{2} \otimes H
	\,,
	\label{eq:had}	
\end{equation}
where $H=\begin{pmatrix} \begin{smallmatrix} 1 & 1 \\ 1 & -1 \end{smallmatrix} \end{pmatrix}/\sqrt{2}$.
Then, $H_3 = M U_H M^\dagger$, with $M$ an unitary $8\times 8$ matrix. This implies the existence of a basis of states 
\begin{align}
	\ket{\psi_i} = \sum_{j=1}^8\ket{\psi_{\diamond}^{(15,2(j-1))}} M^{\dagger}_{ji}\,,
	\label{eq:plane_new_basis}
\end{align}
in the plane $\Pi_{1}$, for which the holonomy of the curve corresponds to $H_3$. For $n=2$ ($R_{\pi \hat{z}}$), the holonomy is trivial, $U=I_{8}$; for $n=3$, it is $U_{H}$, again.

On the other hand, if the last bipyramidal state $\bpy{15}{14}$ is replaced by $\bpy{15}{15}$, we still obtain an anticoherent $8$-plane,
\begin{equation}
	\label{eq:pi_2}
	\Pi_{2}
	=
	\left\{
	\bpy{15}{0},
	\bpy{15}{2},
	...,
	\bpy{15}{12},
	\bpy{15}{15}\right\}\, ,
\end{equation} 
and the resulting set of states now shares a rotational symmetry around the $z$-axis by $\pi$. In this case the holonomy of the curve is given by
\begin{equation}
	\label{eq:U_byp_2}
	U_{T}
	=
	I_{6}\oplus\sigma_{z}\, .	
\end{equation}
This coincides with the diagonal form of the Toffoli gate (\ref{eq:tof}) for $k=8$. As before, there exists a basis change $V$ within the same plane such that, in this new basis the holonomy of the curve corresponds to $T=V U_{T} V^{\dagger}$.
\section{3-qubit universal quantum computing}
A standard technique used to implement quantum error correction, from both experimental and theoretical perspectives, is \textit{quantum coding}. This approach involves constructing a composite quantum system that contains the system of interest $\mathcal{I}$ and an auxiliary subsystem $\mathcal{A}$, known as the \textit{ancilla}. A unitary operation, implemented by the \textit{coding matrix}, is then applied to the composite system and, after a certain time interval, its inverse is applied to decode the information~\cite{Bri:05}.

\begin{figure}[h!]
	\centerline{\includegraphics[width=3.1in]{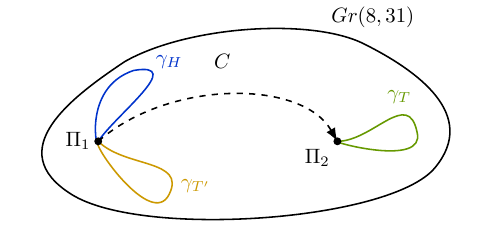}}
	\vspace*{8pt}
	\caption{Schematic of the closed curves $\gamma_{H}$, producing the Hadamard gate on $\Pi_1$, $\gamma_{T}$, producing the Toffoli gate on $\Pi_2$, and $\gamma_{T'}$, defined in~(\ref{eq:tofp_gamma}), producing the Toffoli gate on $\Pi_1$.}
	\label{fig:curves}
\end{figure}

In this section we follow the coding approach to construct a universal set of quantum gates for a $3$-qubit system. As shown above, the Hadamard gate (\ref{eq:had}) can be obtained by rotating the 8-plane spanned by bipyramidal states given in~(\ref{eq:pi_1}) around the $\hat{z}$-axis by $\pi/2$.
Similarly, the Toffoli gate (\ref{eq:tof}) is obtained by rotating the plane in~(\ref{eq:pi_2}) around $\hat{z}$ by $\pi$. Denoting by $C$ the coding matrix that maps the given basis of $\Pi_{1}$ to that of $\Pi_{2}$, consider the closed curve %
\begin{equation}
	\gamma'_{T}:t\in [0,1]
	\mapsto
	(C^{\dagger} R_{\mathbf{n}'(t)} C)\bigl(\Pi_{1}\bigr)
	\, ,
	\label{eq:tofp_gamma}
\end{equation}
based on $\Pi_{1}$, along which the WZ connection vanishes, so that $\gamma_{T}'$ implements, in the TQC scheme, the Toffoli gate (see figure \ref{fig:curves}).
We now consider the $8$-plane $\Pi_{1}$ as the $3$-qubit system $\mathcal{I}$, and the orthogonal complement within the Hilbert space as the ancilla. We use the encoding matrix $C$ in its simplest possible form in the $m$-basis. We now detail the choice of basis and the necessary encoding matrices.

Given the spectral degeneracy of the gate $H_3$, the basis in $\Pi_{1}$ that generates the holonomy associated with $H_3$ is not unique. However, due to the block diagonal structure of both $U_H$ (\ref{eq:U_byp_1}) and $H_3$ (\ref{eq:had}), an immediate change of basis is given by \( M = \mathcal{R}(\frac{\pi}{8})^{\oplus 4} \), where \( \mathcal{R}(\theta) \) is a $2 \times 2$ rotation matrix in \( \mathbb{R}^2 \). That is, the computational basis used to emulate a three-qubit system is given by
\begin{equation}
	\ket{000} = \ket{\psi_1},\quad \ket{001} = \ket{\psi_2},\quad \dots,\quad \ket{111} = \ket{\psi_8} \, ,
	\label{eq:compu_basis}
\end{equation}
where each state in the basis is defined as in (\ref{eq:plane_new_basis}).

Now, since the plane $\Pi_{2}$ is obtained by replacing the state $ \ket{\psi_{\diamond}^{(15,14)}} $ in $\Pi_{1}$ with $\ket{\psi_{\diamond}^{(15,15)}} $, the simplest encoding matrix in the basis of eigenstates of $S_z$, that maps $\Pi_{1}$ to $\Pi_{2}$ is
\begin{equation}
	C_1 = \sigma_x \oplus I_{27} \oplus \sigma_x \, .
	\label{eq:C_mat}
\end{equation}
However, the holonomy induced by the unitary evolution $ C_1^{\dagger} R_{\mathbf{n}'(t)} C_1 $ differs from the Toffoli gate in the $2 \times 2$ block connecting the states $\ket{110}$ and $\ket{111}$. To resolve this discrepancy, we introduce a second encoding matrix that performs a basis change in the subspace
\begin{equation}
	\textup{span}\left\{ \ket{\psi_{\diamond}^{(15,12)}}, \ket{\psi_{\diamond}^{(15,15)}} \right\}
	\subset \Pi_{2} \, .
\end{equation}
The simplest form of this matrix is
\begin{equation}
	C_2 = A \oplus I_{23} \oplus A^J \, ,
	\label{eq:C_mat_2}
\end{equation}
where $A$ is a $4 \times 4$ two-level matrix defined by $A_{1,1} = -A_{4,4} = -\cos\left(\frac{\pi}{8}\right) $, $A_{1,4} = A_{4,1} = \sin\left(\frac{\pi}{8}\right) $, and the rest of its entries as those of the identity matrix. The matrix $A^J$ is obtained by transposing $A$ along its secondary diagonal. The complete encoding matrix $C = C_{2}C_{1}$, written in the $m$-basis, acts non-trivially only on two blocks: those corresponding to the $S_z$ eigenstates with $m = 12, 14, 15$ and $m = -12, -14, -15$, respectively, where the same three-level matrix is applied in both cases. This encoding guarantees that the holonomy generated by the curve $\gamma'_T$, defined in (\ref{eq:tofp_gamma}), matches exactly the Toffoli gate in the computational basis defined in $\Pi_{1}$.

It was shown that in a system of $N$ qubits, the set consisting of all Toffoli gates acting on all tripartite subsystems, together with single-qubit Hadamard gates, forms a universal basis for quantum computation~\cite{Shi.Bot:02,Aha:03}. In contrast, the model under consideration is restricted to emulating a $3$-qubit system as an $8$-dimensional subspace within the Hilbert space of a spin-$15$ system. In this subspace, the gates $T$ and $H_3$ admit implementation under the TQC scheme. Furthermore, the remaining Toffoli and Hadamard gates acting on the first and second qubits are derived from permutations of the computational basis, which are also implemented via additional encoding matrices. Therefore, it is possible to generate a universal set of quantum gates, and hence all required quantum operations for $3$-qubit systems, through rotations and coding techniques, such that the resulting computation is robust against perturbations (within the rotation group) that leave the endpoints of the evolution fixed.
\section{Conclusions}
We used the basic idea of topological quantum computing, together with coding techniques, to implement 3-qubit Toffoli and Hadamard gates. While in the standard TQC protocol quantum gates are implemented by rotations of symmetric spin-planes, in our work, the use of coding matrices allows for more general loops on the Grassmannian, like $\gamma_{T'}$ in (\ref{eq:tofp_gamma}) (see also Fig.~\ref{fig:curves}), which is obtained by conjugation of the rotational curve $\gamma_T$, but which is not, itself, a rotational curve. This opens up the possibility of using a wider class of loops in TQC, leading to the implementation of a wider class of quantum gates. The relevance of anticoherent spin-planes in TQC, and their known exceptional properties in quantum metrology, both suggest that a more systematic study of these geometric objects is now overdue --- we plan on addressing this need in future work. 
\section*{Acknowledgments}

The authors acknowledge partial financial support from the DGAPA-UNAM PAPIIT project IN112224 and the SECIHTI (Mexico) project CBF-2025-I-676. LAM also thanks SECIHTI for a postdoctoral fellowship (CVU 857223).

\bibliographystyle{unsrt}
\bibliography{strings}

\end{document}